\documentclass[12pt,a4paper]{article}
\pdfoutput=1

\usepackage[utf8]{inputenc}

\usepackage{multirow}
\usepackage{amsmath}
\usepackage{color}
\usepackage{amsfonts}
\usepackage{amssymb}
\usepackage{graphicx}
\usepackage{geometry}
\usepackage{amssymb,epsfig,subfigure}
\usepackage{hyperref}
\usepackage{url}
\usepackage{comment}
\usepackage[font=footnotesize]{caption}
\usepackage{slashed}
\usepackage{cite}

\makeatletter
\renewcommand\section{\@startsection {section}{1}{\z@}%
                                 {-3.5ex \@plus -1ex \@minus -.2ex}
                                   {2.3ex \@plus.2ex}%
                                   {\normalfont\large\bfseries}}
\renewcommand\subsection{\@startsection{subsection}{2}{\z@}%
                                   {-3.25ex\@plus -1ex \@minus -.2ex}%
                                     {1.5ex \@plus .2ex}%
                                     {\normalfont\bfseries}}
\renewcommand\subsubsection{\@startsection{subsubsection}{3}{\z@}%
                                   {-3.25ex\@plus -1ex \@minus -.2ex}%
                                     {1.5ex \@plus .2ex}%
                                     {\normalfont\itshape}}
\makeatother

\def\pplogo{\vbox{\kern-\headheight\kern -29pt
\halign{##&##\hfil\cr&{\ppnumber}\cr\rule{0pt}{2.5ex}&\ppdate\cr}}}
\makeatletter
\def\ps@firstpage{\ps@empty \def\@oddhead{\hss\pplogo}%
  \let\@evenhead\@oddhead 
}
\def\maketitle{\par
 \begingroup
 \def\thefootnote{\fnsymbol{footnote}}
 \def\@makefnmark{\hbox{$^{\@thefnmark}$\hss}}
 \if@twocolumn
 \twocolumn[\@maketitle]
 \else \newpage
 \global\@topnum\z@ \@maketitle \fi\thispagestyle{firstpage}\@thanks
 \endgroup
 \setcounter{footnote}{0}
 \let\maketitle\relax
 \let\@maketitle\relax
 \gdef\@thanks{}\gdef\@author{}\gdef\@title{}\let\thanks\relax}
\makeatother

\numberwithin{equation}{section}

\newcommand\eea{\end{eqnarray}}
\newcommand\bea{\begin{eqnarray}}

\def\beq{\begin{equation}}
\def\eeq{\end{equation}}

\newcommand{\be}{\begin{equation}}
\newcommand{\ee}{\end{equation}}
\newcommand{\ba}{\begin{align}}
\newcommand{\ea}{\end{align}}
\newcommand{\bg}{\begin{gather}}
\newcommand{\eg}{\end{gather}}
\newcommand{\bseq}{\begin{subequations}}
\newcommand{\eseq}{\end{subequations}}

\renewcommand{\tanh}{\mathop{\rm th}\nolimits}

\renewcommand{\t}{\tilde}

\newcommand{\mc}{\mathcal}

\newcommand{\dd}[1]{\text{d}#1}

\newcommand{\of}[1]{\left(#1\right)}
\newcommand{\off}[1]{\left[#1\right]}

\newcommand{\coment}[1]{}

\begin{document}
\setcounter{page}0
\def\ppnumber{\vbox{\baselineskip14pt
}}
\def\ppdate{
} \date{}

\author{Nicolás Abate$^{1, 2 }$, Gonzalo Torroba$^{1, 2}$\\
[7mm] \\
{\normalsize \it $^1$Centro At\'omico Bariloche and CONICET}\\
{\normalsize \it $^2$ Instituto Balseiro, UNCuyo and CNEA}\\
{\normalsize \it S.C. de Bariloche, R\'io Negro, R8402AGP, Argentina}\\
}

\bigskip
\title{\bf  Irreversibility of quantum field theory in de Sitter: the C, F and A theorems
\vskip 0.5cm}
\maketitle

\begin{abstract}
We prove the $C$, $F$ and $A$ irreversibility theorems in de Sitter spacetime for quantum field theories that are obtained as renormalization group flows from ultraviolet conformal fixed points. The proof is based on strong subadditivity of the entanglement entropy, de Sitter invariance, and the Markov property of conformal field theory.\end{abstract}
\bigskip

\newpage

\tableofcontents

\vskip 1cm

\section{Introduction}\label{sec:intro}

Understanding quantum field theory (QFT) in cosmological space-times is a fascinating open problem, as well as a necessary step towards formulating the physical laws in our universe. The interaction with a space-time dependent metric introduces phenomena that have no counterpart in flat space-time \cite{Birrell:1982ix}, and raises challenging problems related to cosmological horizons. In flat space, the renormalization group (RG) has provided a foundational framework to describe physics at different length scales in terms of flows of the couplings in the Hamiltonian \cite{Wilson:1973jj}. However, the extension of RG methods to cosmological scenarios is less well developed and requires further exploration.

A key property of the RG in Minkowski space is that it is an irreversible process. This has been established in $d=2,\,3$ and $4$ space-time dimensions, leading to the C, F and A theorems, respectively \cite{Zamolodchikov:1986gt, Casini:2012ei, Komargodski:2011vj, Casini:2017vbe, Hartman:2023qdn}. These results have generated a rich interplay between unitarity and correlator methods, and results from quantum information theory. In this work, our goal  is to establish the irreversibility of the RG for QFT in the simplest and most symmetric cosmological setting, the $d$-dimensional de Sitter universe, dS$_d$. Recent progress in this direction has been achieved by \cite{Loparco:2024ibp, Abate:2024xyb}, who demonstrated the validity of the C-theorem in dS$_2$.

Understanding nonperturbative properties of field theories in de Sitter, obtained as RG flows from conformal fixed points, is important for different reasons. Conformal field theories in de Sitter have been actively studied for their possible relevance in cosmology, for instance for dynamical phase transitions or as strongly coupled sectors during inflation \cite{Marolf:2010tg}. Theories with light and/or derivatively coupled scalar fields are directly relevant for inflation, and could be completed into interacting theories -- see e.g. \cite{Maldacena:2002vr, Chu:2016pea, Pimentel:2025rds} and references therein. Furthermore, conceptually, de Sitter gives rise to new phenomena, related to the absence of a global energy, to the existence of a cosmological horizon and to strong infrared effects, all of which are absent from flat space.

In this work, we prove that the information-theoretic approach of \cite{Casini:2017vbe} can be generalized from Minkowski to de Sitter. Using strong subadditivity (SSA) of the entanglement entropy and de Sitter invariance, we derive the irreversibility formula
\be\label{eq:bSSA}
R^2 \Delta S''(R)-(d-3) R\, \Delta S'(R) \le 0\,,
\ee
where $\Delta S$ is the difference of entanglement entropies between the UV fixed point and the QFT along the RG flow, evaluated on a causal diamond of radius $R$ inside the static patch of dS$_d$. The result has the same form as its Minkowski counterpart, while heavily relying on properties of dS$_d$. For $d=2$ our formula recovers the C-theorem;  for $d=3$ it implies the F-theorem in $dS$; and for $d=4$ it gives the A-theorem. These results provide new insights on nonperturbative QFT in de Sitter space, highlighting the utility of quantum information theory in cosmology.

The work is organized as follows. In section \ref{sec:setup} we introduce the kind of theories we will analyze and present the general approach to study RG flows using entanglement. In section \ref{sec:markov} we establish the Markov property for CFTs in dS. This will be crucial in the derivation of the irreversibility formula, which we later show in section \ref{sec:irrev}. In section \ref{sec:RGflow} we comment on the consequences of the irreversibility formula for RG flows in de Sitter. Finally, we end with a general discussion of our results and some possible future directions in section \ref{sec:concl}.

\section{Setup}\label{sec:setup}

In this section we introduce the basic geometry and define the entanglement entropy for spherical regions. The derivation of the main irreversibility inequality is deferred to Sec.~\ref{sec:irrev}, and its physical consequences are discussed in Sec.~\ref{sec:RGflow}.

\subsection{Assumptions}

We consider unitary and de Sitter invariant quantum field theories (QFTs) in $d$-dimensional de Sitter space-time, dS$_d$. In global conformal coordinates, the metric is
\be\label{eq:metric1}
\dd s_d^2= \frac{\ell^2}{\cos^2T}\,\left(-\dd T^2+\dd \theta^2 + \sin^2 \theta\,\dd\Omega_{d-2}^2 \right)\,,
\ee
where $\ell$ is the dS radius, and the coordinate ranges $-\pi/2 \le T \le \pi/2$, $0 \le \theta \le \pi$ cover the full space-time. We focus on the space of QFTs that are obtained by perturbing a conformal field theory (CFT) by relevant scalar primary operators,
\be\label{eq:Sqft}
\mc S_\mathrm{QFT}= \mc S_\mathrm{CFT}+ \int \dd^dx\,\sqrt{-g}\,\lambda_I \phi_I\,,
\ee
where the short distance scaling dimension of $\phi_I$ at the fixed point satisfies $\Delta_I<d$. The CFT acts as the ultraviolet fixed point for the QFT. We will work with the dS invariant Bunch-Davies or euclidean vacuum \cite{Chernikov:1968zm, Bunch:1978yq, Mottola:1984ar, Allen:1985ux}. The deformation (\ref{eq:Sqft}) triggers a nontrivial renormalization group flow; correlation functions in the euclidean vacuum at a typical distance scale $\Delta x$ generally depend on $\Delta x/\ell$ and $\lambda_I \ell^{d-\Delta_I}$.

\subsection{Entanglement in dS}

\begin{figure}[h]
    \centering
    \includegraphics[width=0.5\textwidth]{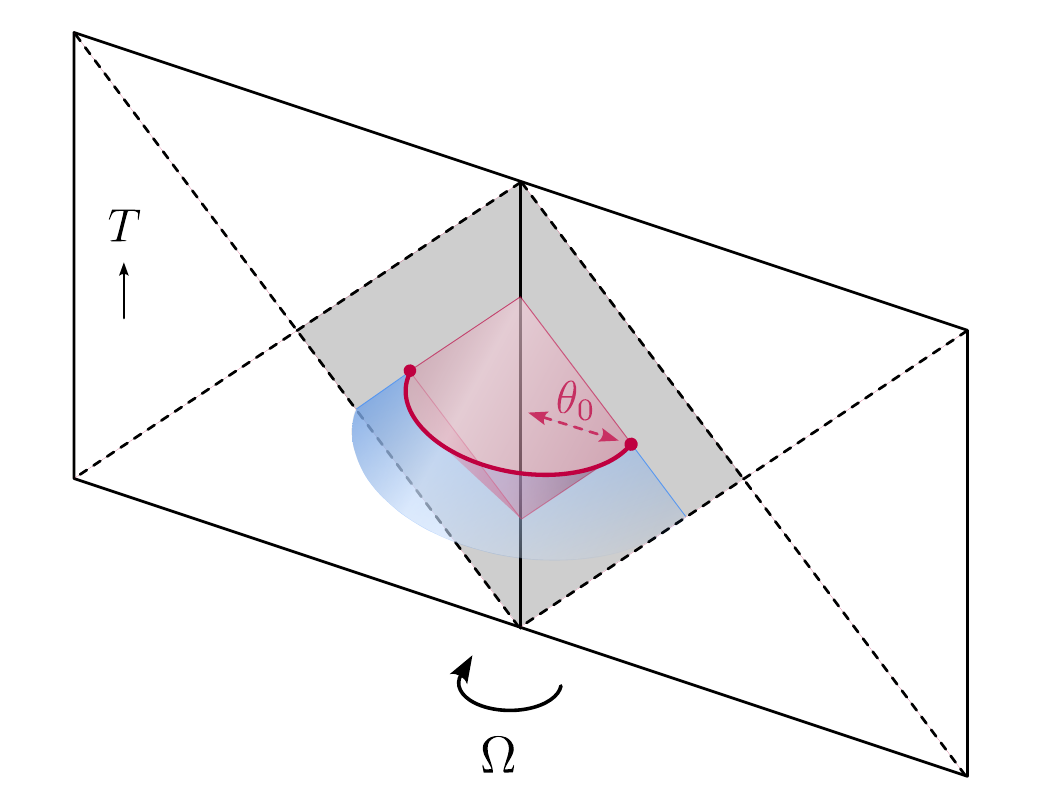}
    \caption{In pink we show the causal diamond associated with the sphere of radius $R=\ell\sin\theta_0$ at $T=0$ over the (unwrapped) Penrose diagram. Its future boundary coincides with the null cone $T=\theta_0-\theta$, which is displayed in blue. The grey diamond is the static patch.}
    \label{fig:naive_cone}
\end{figure}

We will use the entanglement entropy to determine nonperturbative properties of RG flows in dS$_d$. Previous works on entanglement entropy in de Sitter include \cite{Maldacena:2012xp, Ben-Ami:2015zsa, Boutivas:2024lts}. Let us consider the vacuum entanglement entropy (EE) for a sphere of radius $R= \ell \sin \theta_0$
at $T=0$ (the neck of dS), which we show on the Penrose diagram in Figure \ref{fig:naive_cone}. The spatial entangling region is $0\le \theta \le \theta_0$. The future boundary of the associated causal diamond in conformal coordinates is simply the null cone $T=\theta_0-\theta$. 

When $\theta_0=\pi/2$, the entangling radius becomes $R=\ell$; the entangling region covers the full static patch, and the EE becomes the thermal dS entropy. This means that the density matrix becomes $\rho=e^{- H}$, where the modular Hamiltonian $H$ coincides with the physical static patch Hamiltonian (the generator of time translations). Since we are calculating the entanglement entropy with respect to a global pure state (the euclidean vacuum at $T=0$), the entropy of a region is equal to the entropy of its complement, which in this case gives $S(\theta_0) = S(\pi-\theta_0)$. So the EE attains its maximum at $\theta_0=\pi/2$, and for $\theta_0>\pi/2$ the EE coincides with that of a sphere inside the complementary static patch.

The EE for a QFT in dS is a divergent quantity; in general, it is a complicated function of the sphere radius $R$, the dS radius $\ell$, a short distance cutoff $\epsilon$, and the relevant couplings $\lambda_I$ in (\ref{eq:Sqft}). We will be interested in the finite entropy difference
\be\label{eq:DS}
\Delta S(R) = S_\textrm{QFT}(R)-S_\textrm{UV}(R)\,,
\ee
where $S_\textrm{UV}(R)$ is the entropy of the UV fixed point (plus appropriate local counterterms). We will analyze this subtraction procedure in more detail below.

\subsection{Irreversibility}

In order to characterize the irreversibility of RG flows in dS, let us now describe the entropy for a CFT in dS. As we discuss in Sec. \ref{sec:markov}, the existence of a conformal transformation between the causal diamond of a sphere in dS and that of a sphere in flat space, implies that the structure of the EE is the same as that in flat space:
\be\label{eq:Scft}
S(R)=\mu_{d-2}\,R^{d-2}+\mu_{d-4}\, R^{d-4}+\ldots + \left\lbrace \begin{array}{ll} (-)^{\frac{d-2}{2}} 4\,A\, \log(R/\epsilon)\,& d\;  \textrm{even}\,\\ (-)^{\frac{d-1}{2}} F\,& d\,\,\textrm{odd} \, \end{array}\right.
\ee
where the coefficients are proportional to inverse powers of the cutoff, $\mu_{k}\sim \epsilon^{-k}$. Our focus is on the quantities $A$ and $F$, which are finite and intrinsic to the fixed point. They coincide with the corresponding $A$ and $F$ terms of the EE in flat space, and they can be computed as the universal terms of the free energy for a CFT on a sphere. $A$ is also the Euler anomaly in even $d$, which for $d=2$ is proportional to the central charge $C$ of the CFT.

For a QFT in flat space, the UV fixed point can be probed in terms of the EE by taking $R \to 0$, and the IR fixed point can be studied taking $R \to \infty$. The irreversibility of the RG involves establishing that $A$ or $F$ decrease between the UV and IR fixed points,
\be\label{eq:irrevMink}
A_\mathrm{UV} > A_\mathrm{IR}\;,\;F_\mathrm{UV}> F_\mathrm{IR}\,,
\ee
where the UV and IR RG charges are intrinsic to the fixed points, namely they are finite and independent of the RG flow. However, for a QFT in dS this notion needs to be generalized, because there is a maximum entangling distance scale $R=\ell$, corresponding to the largest causally accessible region to an observer. Physically, the de Sitter temperature of the static patch limits the smallest energies that can be probed.

To extend the concept of irreversibility of the RG to dS our strategy will be to use EE inequalities. Building on the results of Secs. \ref{sec:markov} and \ref{sec:irrev}, we will demonstrate that (\ref{eq:DS}) satisfies the irreversibility formula (\ref{eq:bSSA}). Then we will define a running A-function $\Delta A(R) = A(R) - A_\textrm{UV}$ (as well as analogous functions $\Delta C$ and $\Delta F$), and prove that these functions have to be negative for all unitary and de Sitter invariant QFTs. This construction provides a natural notion of irreversibility in dS in terms of finite physical quantities. 

At the maximum value $R=\ell$, $\Delta S(\ell)$ coincides with a thermal entropy difference, analogous to the Rindler limit $R \to \infty$ in flat space. Furthermore, taking the limit $\ell \to \infty$ recovers the results for Minkowski space. Unlike (\ref{eq:irrevMink}), this notion of irreversibility at the largest possible infrared scale gives
\be\label{eq:irrevdS}
A_\mathrm{{UV}} > A(R=\ell)\;,\;F_{\mathrm{UV}}> F(R=\ell)\,,
\ee
and the RG charges at $R=\ell$ are not intrinsic to an IR fixed point. Instead, they can depend on the details of the RG flow. This is conceptually different from the irreversibility notion (\ref{eq:irrevMink}), and is a direct consequence of the finite temperature or maximum finite causally accessible size in de Sitter.

We will now establish the key quantum-information ingredients needed to prove these results, namely the Markov property of the CFT vacuum in de Sitter and infinitesimal deformations along a dS light-cone.

\section{Entanglement and Markov property in dS}\label{sec:markov}

The derivation of the irreversibility formula for QFTs in de Sitter will rely on three ingredients: dS invariance, strong subadditivity (SSA) of the EE, and the Markov property for the CFT vacuum. This last property was proved in flat space in \cite{Casini:2017roe}, and the goal of this section is to establish it in de Sitter. In order to do so, we will exploit the conformal invariance to derive the result for dS by mapping from flat-space.

Our analysis will require generalizing the spherical entangling regions to more general regions with boundary on a common light-cone inside the static patch. We will take this light-cone to correspond to the past light-cone from the origin. In the proof of the irreversibility formula in Sec. \ref{sec:irrev}, these boundaries will arise from applying a de Sitter boost to a spatial sphere. 

\subsection{Markov property in dS}

We now prove the Markov property for the CFT vacuum in de Sitter. The argument proceeds in two steps: (i) using conformal invariance, we show that the CFT entanglement entropy for cuts on the null cone is a local functional of the cut (up to a possible anomaly term); and (ii) we verify that the anomaly term is itself local, which implies that strong subadditivity is saturated for the CFT. This subsection is devoted to step (i), while step (ii) is taken in the next subsection.

\begin{figure}[h]
    \centering  \includegraphics[width=0.3\textwidth]{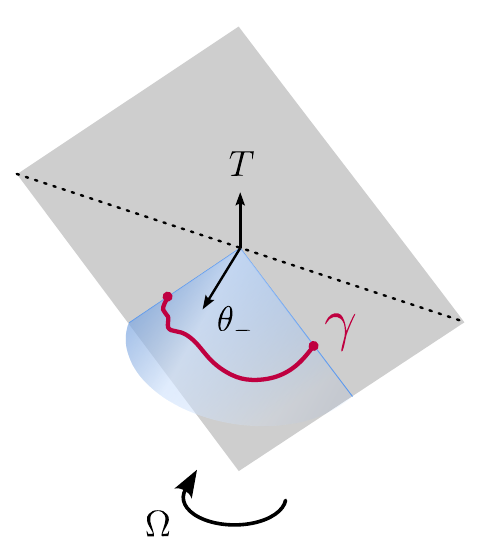}
    \caption{In order to arrive at our irreversibility formula, we will use entangling regions with boundary given by some curve $\gamma(\hat x)$ on the past null cone passing through the origin defined by the equations $\theta_+=\theta+T=0$ and $\theta_-=\theta-T=2\gamma$, always restricted to be inside the static patch.}
    \label{fig:right_cone}
\end{figure}

Consider a general curve on the past null cone from the origin inside the static patch of dS, as shown in Figure \ref{fig:right_cone}, parametrized by
\be\label{eq:def_gamma}
T(\hat{x})=-\theta(\hat{x})=-\gamma(\hat{x})\,,
\ee
which defines the boundary of some entangling region. Here $\hat x^i$ is a unit vector describing the sphere $S^{d-2}$. The metric (\ref{eq:metric1}) restricted to this null boundary describes a $(d-2)$-dimensional sphere with varying radius
\be\label{eq:dgamma1}
\dd s^2\Big|_\gamma=\ell^2\tan^2\gamma(\hat x)\,\dd\Omega_{d-2}^2\,.
\ee

We want to determine the functional dependence of the EE for the CFT on a region with boundary $\gamma(\hat x)$ on the null cone inside the static patch of dS. This entropy is divergent, and we assume that it is regularized by a dS invariant cutoff $\epsilon$ (for instance using the mutual information \cite{Casini:2015woa}). Since the causal diamond is inside the static patch, we can compose the transformations of \cite{Candelas:1978gf, casini:2011kv} in order to conformally map this region onto a region with boundary on the null plane in flat space, $x^0=-x^1=-\t \gamma(y)$ (where $y$ are the remaining $d-2$ coordinates, and $\t \gamma$ is a local function of $\gamma(\hat x)$). If the CFT in this flat space has a Lorentz-invariant regularization, then the vacuum EE is independent of $\t \gamma$ \cite{Casini:2018kzx}. A simple way to see this is that $S[\t \gamma]$ is Lorentz invariant; under a Lorentz boost, $S[\t \gamma]= S[\lambda \t \gamma]$, and taking $\lambda \to 0$ we can transform the curved boundary into the Rindler edge.

However, in general the dS invariant UV cutoff $\epsilon$ does not map to a Lorentz invariant cutoff in the flat space theory. This means that the EE in dS can only depend on the boundary curve at short distances via the UV cutoff as it happens in flat space. In other words, the CFT entropy in dS will be independent of the entangling regions when taking variations at non-coincident points,
\be\label{eq:derivS}
\frac{\delta^{(n)} S_{\textrm{CFT}}[\gamma]}{\delta \gamma(\hat x_1) \ldots \delta \gamma(\hat x_n)}=0\;,\;\textrm{for}\;\;\hat x_1 \neq \ldots \neq \hat x_n\,,
\ee
but due to the presence of short distance divergences, the functional variations at coincident points can be nonzero (and are in general divergent).

\subsection{Structure of the CFT EE}

The result (\ref{eq:derivS}) implies that the entropy should be a local quantity of geometric boundary objects, up to a possibly non-local term whose variations at separated points have to vanish:
\be\label{eq:localS}
S_\textrm{CFT}[\gamma]= \int \dd\Omega \,\sqrt{\hat g}\,L(\ell \tan \gamma/\epsilon, \hat g, \partial, \ldots)+ F_\textrm{non-loc}\,,
\ee
where $\hat g$ is the unit radius sphere metric. As discussed in \cite{Casini:2018kzx}, $F_\textrm{non-loc}$ can be either a constant, or a bilocal (anomaly) action. 

$F_\textrm{non-loc}$ gives the origin of the universal entropy term in odd and even space-time dimensions respectively, the later corresponding to the Weyl anomaly. We give more details on the expression \eqref{eq:localS} in App. \ref{appendix:K}. To illustrate the structure of the local functional, we now display the explicit terms for a constant-angle sphere in the spacetime dimensions relevant to the C, F, and A theorems:
\begin{align}
S_\mathrm{CFT}^{(d=2)}(\gamma_0)&=\frac{C}{3}\log\of{\frac{\ell\tan\gamma_0}{\epsilon}}\,,\label{eq:S2d}\\
S_\mathrm{CFT}^{(d=3)}(\gamma_0)&=\alpha_0\,\frac{\ell \tan\gamma_0}{\epsilon}-F\,,\label{eq:S3d}\\
S_\mathrm{CFT}^{(d=4)}(\gamma_0)&=\alpha_0\,\frac{\ell^2\tan^2\gamma_0}{\epsilon^2}-4A\log\of{\frac{\ell\tan\gamma_0}{\epsilon}}\,.\label{eq:S4d}
\end{align}
Note that the three expressions are consistent with \eqref{eq:Scft} and reduce to the well-known flat-space formulae in the limit $\gamma_0\ll1$.

In particular, \eqref{eq:derivS} implies that the EE in dS satisfies
\be\label{eq:markov_S0}
\frac{\delta^{(2)} S_\textrm{CFT}[\gamma]}{\delta \gamma (\hat x_1)  \delta \gamma(\hat x_2)}=0\;,\,\hat x_1 \neq \hat x_2 
\ee
under null deformations of the boundary region. This infinitesimal equality is equivalent to the formula
\be\label{eq:markov_S}
S_\mathrm{CFT}^A+S_\mathrm{CFT}^B-S_\mathrm{CFT}^{A\cup B}-S_\mathrm{CFT}^{A\cap B}=0\,,
\ee
where $A$ and $B$ are finite regions with boundary on the null cone.\footnote{The left hand side of (\ref{eq:markov_S0}) is the infinitesimal version of the the left hand side of (\ref{eq:markov_S}). To see this, attach infinitesimal bumps $B_1, B_2$ to a region $W$, i.e. $A= W \cup B_1$, $B= W \cup B_2$. Then the left hand side of (\ref{eq:markov_S}) becomes a difference of two functional first derivatives, giving rise to the functional second derivative in (\ref{eq:markov_S0}). }
In other words, strong subadditivity is saturated for the EE of a CFT with boundaries on the null cone in de Sitter. This is the Markov property for the CFT vacuum, which was proved in Minkowski in \cite{Casini:2017roe}, and we now see that it is also valid in de Sitter. Eq. (\ref{eq:markov_S0}) will be important for the derivation of the irreversibility formula in the next section.

\section{Irreversibility formula}\label{sec:irrev}

Our goal now is to prove (\ref{eq:bSSA}). The strategy is to use the dS isometries, to deform the entangling region and derive a second order relation for the entanglement entropy. This, combined with SSA, will lead to the irreversibility formula. This approach was used in flat space in \cite{Casini:2023kyj}.

\subsection{Infinitesimal deformations and dS invariance}

Consider a spatial sphere given by (\ref{eq:def_gamma}) with constant $\gamma=\gamma_0$.
The intrinsic radius is $R=\ell\,\tan\gamma_0$.
Under a small dS boost of the curve, we can expand the EE in functional derivatives, and since the vacuum state is dS invariant, we arrive at
\begin{align}\label{eq:invnew}
0=\int_{\Omega_1}\,S_1\,\delta\gamma_1+\frac{1}{2}\int_{\Omega_1,\Omega_2}S_{12}\,\delta\gamma_1\delta\gamma_2+\dots\,,
\end{align}
which gives, to every order in the boost parameter, an infinite hierarchy of equalities between derivatives of the EE. Here $\delta\gamma_i=\delta\gamma(\hat x_i)$, we have denoted
\be
S_1(\hat x_1) = \frac{\delta S}{\delta \gamma(\hat x_1)}\Big|_{\gamma_0},\;S_{12}(\hat x_1, \hat x_2) =\frac{\delta^2 S}{\delta \gamma(\hat x_1)\delta \gamma(\hat x_2)}\Big|_{\gamma_0}\,,
\ee
and the angular integration $\int_\Omega= \int \dd^{d-2}\Omega$ is normalized such that $\int_\Omega\,1=1$.
Note that
\begin{align}
\int_{\Omega_1} S_1 &= \frac{\ell}{\cos^2\gamma_0}\,S'(R)\,,\\
\int_{\Omega_1,\Omega_2} S_{12} &= \frac{2\ell\tan\gamma_0}{\cos^2\gamma_0}\,S'(R)+\frac{\ell^2}{\cos^4\gamma_0}\,S''(R)\,, \nonumber
\end{align}
which follows from changing variables from $\gamma$ to $R=\ell \tan \gamma$.
Also, rotational invariance imposes some restrictions: $S_1$  has to be a constant and thus $S_1(\hat x_1)=\ell\sec^2\gamma_0\,S'(R)$, and $S_{12}(\hat x_1,\hat x_2) = S_{12}(\hat x_1 \cdot \hat x_2)$.

We will now consider a specific type of infinitesimal deformation that arises from a dS boost.
It is convenient to perform this analysis in terms of the embedding of dS$_d$ in a $(d+1)-$Minkowski spacetime as a hyperboloid of radius $\ell$. Namely
\be\label{eq:dS_embedding}
-(X^0)^2+(X^d)^2+(X^i)^2=\ell^2\,,
\ee
where $i=1,\dots,d-1$. Different parametrizations of the hyperboloid yield different coordinate systems on dS. Choosing
\be
X^0=\ell\tan T,\ X^d=\ell\frac{\cos\theta}{\cos T},\ X^i=\ell\frac{\sin\theta}{\cos T}\hat{x}^i\,,
\ee
with $|T|\leq\pi/2$, $0\leq\theta\leq\pi$ and $\hat{x}^i$ describing a vector on the sphere $S^{d-2}$, gives the global conformal coordinates with metric \eqref{eq:metric1}.
The connected isometry group of dS$_d$ can thus be obtained from the embedding, as the restriction of the Lorentz generators which preserve the hyperboloid. These are the generators $J^{MN}$ of the group $SO(1,d)$ restricted to \eqref{eq:dS_embedding}. Naturally, the Bunch-Davies vacuum is invariant under these isometries. In particular, the boosts $J^{0i}$ preserve the light-cone through the origin $T=-\theta$, and we will be interested on using these generators to deform spatial spheres with boundary on the past cone. 

Consider a boost $J_{0i}\hat{n}^i$ of parameter 
$\eta$ along one direction $\hat{n}^i$ acting on some spatial sphere given by $T=-\theta=-\gamma_0$. Its effect on a point $X^M$ on the sphere gives the relation
\be\label{eq:boosted1}
X^0(\eta)=-\frac{\ell\tan\gamma_0}{\cosh\eta}+\tanh\eta\,X_n(\eta)\,,
\ee
where $X_n=X_i\hat{n}^i$. This parametrizes a boosted sphere with boundary $\gamma_\eta(\hat{x})$ on the cone. Writing $X^0(\eta)=-\ell\tan\gamma_\eta$,  $X_n(\eta)=\ell\tan\gamma_\eta\,(\hat n\cdot \hat x)$ and plugging back into \eqref{eq:boosted1}, we can solve for $\gamma_\eta$:
\be
\tan\gamma_\eta(\hat{x})=\frac{\tan\gamma_0}{\cosh\eta+\sinh\eta(\hat{n}\cdot\hat{x})}\,.
\ee
So taking $\eta\ll1$, we arrive at the formula for infinitesimal deformations
\be\label{eq:dgamma}
\gamma_\eta=\gamma_0+\delta\gamma=\gamma_0-\eta\,\cos\gamma_0\sin\gamma_0\of{\hat n\cdot\hat x}+\eta^2\cos\gamma_0\sin\gamma_0\of{\of{\hat n\cdot \hat x}^2\cos^2\gamma_0-\frac{1}{2}}+\dots
\ee

Plugging (\ref{eq:dgamma}) into \eqref{eq:invnew}, to first order on $\eta$ one obtains an equality 
which is automatically satisfied by rotational invariance. Meanwhile, at second order in $\eta$ we have two contributions: one from the first term in (\ref{eq:invnew}) keeping $\delta \gamma \sim \eta^2$, and one from the second term keeping $\delta \gamma \sim \eta$. The result is
\be\label{eq:constr1}
0=\int_{\Omega_1}\of{(\hat n \cdot \hat x_1)^2\cos^2\gamma_0-\frac{1}{2}}S_1
+\frac{\cos\gamma_0\sin\gamma_0}{2} \int_{\Omega_1,\Omega_2}(\hat n \cdot \hat x_1)(\hat n \cdot \hat x_2) S_{12}(\hat x_1 \cdot \hat x_2)\,.
\ee
Rotational invariance allows to replace products such as $\hat x_1^i \hat x_2^j \to \delta^{ij}/(d-1)$ inside integrals. Then (\ref{eq:constr1}) is equivalent to
\be\label{eq:equality2}
-\of{\frac{d-1}{\cos^2\gamma_0}-2}\ell S'(R) +\cos\gamma_0\sin\gamma_0\, \alpha(R)=0\,,
\ee
where we have defined
\be
\alpha(R)=\int_{\Omega_1,\Omega_2} \hat x_1\cdot \hat x_2\,S_{12}(\hat x_1 \cdot \hat x_2)\,.
\ee

\subsection{Proof of irreversibility formula}

Now we use strong subadditivity, which says that $S_{12} \le 0$ when the points do not coincide. This is not sufficient though, because the integral also receives a contribution from $\hat x_1 = \hat x_2$, and in this case the sign of $S_{12}$ is not determined. This is the step in which we use the Markov property. We construct the entropy difference
\be\label{eq:DS1}
\Delta S[\gamma] = S_\textrm{QFT}[\gamma]-S_\textrm{UV}[\gamma]\,,
\ee
where $S_\textrm{UV}$ is the entropy of the UV fixed point plus possible UV divergences induced by the relevant perturbations. This term is Markovian (the UV divergent terms are local on the entangling surface), and hence $\Delta S_{12} \le 0$, and it vanishes at coincident points because the UV behavior of both terms is the same.\footnote{The subtraction of divergences can be implemented in terms of local counterterms in the field theory action, corresponding to a local gravitational action. The simplest example is the renormalization of the area term in the entropy, which is equivalent to a renormalization of Newton's constant. See e.g. \cite{Cooperman:2013iqr, Casini:2014yca}.}
Therefore,
\be
\int_{\Omega_1,\Omega_2} (1- \hat x_1 \cdot \hat x_2)\Delta S_{12}(\hat x_1 \cdot \hat x_2)\le 0\,.
\ee
Finally, recognizing that the second term here is $\alpha(R)$, and replacing it in terms of $S'(R)$ using (\ref{eq:equality2}), we arrive at our main result
\be\label{eq:irrev2}
R \Delta S''(R)-(d-3) \Delta S'(R) \le 0\,.
\ee

\subsection{Application to the static and inflationary patches}

The above formula was written for $R= \ell \tan \theta_0$ and for variations along the light-cone, namely $T=-\theta$. Therefore, even though it is the same as the analog formula in flat space-time, it leads to results that are conceptually different due to the space-time dependent metric. In order to see this, we will now write it in terms of the geodesic radius $\rho$ of the spherical region, namely half of the geodesic distance between antipodal points of the boundary of the entangling region. We will then apply the result to two cases of interest: the static patch and the inflationary coordinates.

The geodesic distance can be easily evaluated in the embedding space for dS$_d$,
\be
-(X^0)^2+(X^d)^2+ (X^i)^2=\ell^2\;,\;i=1,\,\ldots, \,d-1\,,
\ee
where $\ell$ is the dS radius, and the global conformal coordinates correspond to
\be\label{eq:embedding-conf}
X^0=\ell\,\tan T\;,\;X^d=\ell\, \frac{\cos \theta}{\cos T}\;,\;X^i=\ell\,\frac{\sin \theta}{\cos T}\, \hat x^i\;,\;i=1,\,\ldots, \,d-1\,,
\ee
Denoting by $\eta_{AB}$ the Minkowski metric of the embedding space, and the inner product between two points $x$, $y$ on the hyperboloid by
\be
Z(x,y)=\frac{1}{\ell^2} \eta_{AB} X^A(x) X^B(y)\,,
\ee
the geodesic distance $D(x,y)$ between the two points is given by the standard formula
\be
\cos \frac{D(x,y)}{\ell}=Z(x,y)\,.
\ee

We can further relate the geodesic distance with the intrinsic radius $R$ of the spherical entangling region. In our first case of interest, we consider a sphere of coordinate radius $\theta_0$ sitting at the neck of dS ($T=0$) with antipodal points
\be 
X_{\mathrm{neck}}^0=0\;,\;X_{\mathrm{neck}}^d=\ell\cos\theta_0\;,\;X_{\mathrm{neck}}^i=\ell\sin\theta_0\hat n^i
\ee
and
\be 
Y_{\mathrm{neck}}^0=0\;,\; Y_{\mathrm{neck}}^d=\ell\cos\theta_0\;,\; Y_{\mathrm{neck}}^i=-\ell\sin\theta_0\hat n^i\,,
\ee
that is we simply reflect the unit vector on the $S^{d-2}$ sphere, namely $\hat n^i\to-\hat n^i$. Defining the geodesic radius of the sphere as
\be 
\rho=D(x,y)/2\,,
\ee
a short calculation gives
\be\label{eq:Randrho_neck} 
R=\ell\,\sin\theta_0=\ell\,\sin\frac{\rho}{\ell}\,.
\ee
Instead, taking the surface on the null cone $T=-\theta=-\gamma_0$ the antipodal points are
\be
X^0_{\mathrm{cone}}=- \ell \,\tan \gamma_0\;,\;X^d_{\mathrm{cone}}=\ell\;,\;X^i_{\mathrm{cone}}=- \ell \,\tan \gamma_0\, \hat n^i
\ee
and
\be
Y^0_{\mathrm{cone}}=- \ell \,\tan \gamma_0\;,\;Y^d_{\mathrm{cone}}=\ell\;,\;Y^i_{\mathrm{cone}}= \ell \,\tan \gamma_0 \,\hat n^i\,.
\ee
Again one gets
\be\label{eq:Randrho}
R= \ell \,\tan \gamma_0 = \ell\,\sin \frac{\rho}{\ell}\,.
\ee

By de Sitter invariance, we can then write the entanglement entropy as a function of the geodesic radius of the sphere. The irreversibility inequality becomes
\be\label{eq:irrevgeo}
\ell\,\tan \frac{\rho}{\ell}\, \Delta S''(\rho) - \left(d-3- \tan^2 \frac{\rho}{\ell}\right) \Delta S'(\rho) \le 0\,.
\ee
In this form, the effects from the curved space-time become manifest. First, for small radius $\rho \ll \ell$, we recover the Minkowski result,
\be
\rho \, \Delta S''(\rho) - (d-3)  \Delta S'(\rho) \lesssim 0\,.
\ee
This is a consistency check, since at small distances the dynamics should proceed as in flat space-time. On the other hand, for larger $\rho$ we find strong differences. In particular, near the largest possible radius $ \rho_\textrm{max}=\frac{\pi}{2} \ell$ (the limit where the causal diamond covers the full static patch), the tangent function diverges and
\be
\ell\,\Delta S''(\rho)+\tan \frac{\rho}{\ell}\, \Delta S'(\rho) \lesssim 0\,.
\ee
We note that as $\rho/\ell \to \pi/2$, by symmetry $\Delta S' \to 0$, and this inequality is automatically satisfied.

Let us now specialize (\ref{eq:irrevgeo}) to the static and inflationary patches. The dS static patch metric is
\be
\dd s^2=- \of{1-r^2/\ell^2}\dd t^2+ \frac{\dd r^2}{1-r^2/\ell^2}+r^2\, \dd\Omega_{d-2}^2\,.
\ee
We consider a spherical entangling region 
\be\label{eq:spherestatic}
t=0\;,\;0 \le r \le R\,.
\ee 
In global coordinates, this maps to a region at $T=0$ and $0\le \theta \le \theta_0$ with $\ell \sin \theta_0=R$.
Since the coordinate $r$ determines the area of the $S^{d-2}$ sphere, we see that the entangling surface $r= R$ coincides with $\ell \tan \theta_0$ in the previous subsection. Therefore, in terms of this static patch coordinate, the inequality is the same as (\ref{eq:irrev2}). We can also express the inequality in terms of the sphere geodesic radius $\rho$; see Eq. (\ref{eq:irrevgeo}). In the static patch, the geodesic radius of the spherical region (\ref{eq:spherestatic}) is
\be
\rho= \int_0^R \frac{\dd r}{\sqrt{1-r^2/\ell^2}}= \ell \arcsin \frac{R}{\ell}\,,
\ee
in agreement with (\ref{eq:Randrho_neck}).

Finally, consider the inflationary or Poincaré patch,
\be
\dd s^2 =\frac{\ell^2}{\eta^2}\of{-\dd\eta^2 + \dd \bar x^2}=\frac{\ell^2}{\eta^2}\of{-\dd\eta^2 + \dd  \xi^2+\xi^2 \dd\Omega_{d-2}^2}\,,
\ee
with $\eta \to -\infty $ and $\eta=0$ the asymptotic past and future infinities, respectively. We take a spherical entangling region
\be
\eta=\eta_0\;,\;0\le \xi \le \xi_0\,.
\ee
Note that this Cauchy surface at $\eta=\eta_0$ does not correspond to a constant time slice in the global or static patch coordinates. Nevertheless, the inequality only requires the invariant geodesic distance, so this is not an issue. By comparing the proper area of the sphere here with the previous cases we read off
\be
R= \ell\, \frac{\xi_0}{|\eta_0|}\,.
\ee
If we change the radius of the sphere keeping $\eta_0$ fixed, we have $\dd R=\ell/|\eta_0|\,\dd \xi_0$, and hence the irreversibility inequality in the inflationary patch becomes
\be\label{eq:irrevinflationary}
\xi_0\, \Delta S''(\xi_0)-(d-3) \Delta S'(\xi_0) \le 0\,,\quad\eta=\eta_0\,,\quad \xi_0 \le |\eta_0|\,.
\ee
This is the same as the result in flat space-time, so we conclude that the overall scale factor $\ell^2/\eta^2$ drops out from the inequality. However, unlike flat space-time, here we have the restriction $\xi_0 \le |\eta_0|$ or $R \le \ell$ (the size of the largest causally accessible region).

\section{Irreversibility of the renormalization group in de Sitter}\label{sec:RGflow}

Let us now show how our irreversibility formula \eqref{eq:irrev2} constrains RG flows in de Sitter in different dimensions. 

\begin{itemize}
\item For $d=2$, (\ref{eq:irrev2}) shows that the quantity
\be 
\Delta C(R) = 3R \,\Delta S'(R)\,,
\ee 
montonically decreases along the flow, establishing the C-theorem in dS$_2$. By \eqref{eq:S2d} we can write $\Delta C(R)=C(R)-C_\mathrm{UV}$, where $C(R)=3R\, S'_\mathrm{QFT}(R)$, and \eqref{eq:irrev2} then implies that the $C$ function is bounded by the UV central charge at every intermediate step of the flow. Evaluating it at the maximum causally accesible length $R=\ell$, we have
\be
C(\ell)\leq C_\mathrm{UV}\,.
\ee
This entropic theorem was recently proved in \cite{Abate:2024xyb}.

\item Similarly, in $d=3$ the inequality \eqref{eq:irrev2} implies the monotonic decrease of 
\be 
\Delta F(R)=F(R)-F_\mathrm{UV}=R\,\Delta S'(R)- \Delta S(R)\,.
\ee
This is the F-theorem in dS$_3$. In particular, at $R=\ell$ this implies
\be 
F(\ell)\leq F_\mathrm{UV}\,.
\ee

\item In $d=4$, \eqref{eq:irrev2} establishes that
\be 
\Delta A(R)=A(R)-A_\mathrm{UV}=\tfrac{1}{8}\of{R^2\Delta S''(R)-R\Delta S'(R)},
\ee
is negative, giving the A-theorem in dS$_4$. At $R=\ell$, we also have
\be 
A(\ell)\leq A_\mathrm{UV}\,.
\ee

\item Finally, for $d \geq 5$, the irreversibility formula implies $\Delta \mu_{d-2} \le 0$ and $\Delta \mu_{d-4} \ge 0$, but it is not sufficient to constrain the change of the universal coefficients in the EE. 
\end{itemize}

We note that $\Delta C$, $\Delta F$ and $\Delta A$ vanish in the limit $R \to 0$, and they are defined directly in terms of derivatives or integrals of (\ref{eq:irrev2}). Therefore they are independent of the renormalization procedure defined in (\ref{eq:DS1}). These quantities take the same form as their Minkowski counterparts. However, a key difference is that in flat space one can take $R \to \infty$, and these quantities become differences of IR and UV RG charges, while in dS the largest radius is restricted by the static patch size. Nevertheless, this generalized notion of irreversibility in de Sitter reduces to that of flat space in the limit of large dS radius. Additionally, in both de Sitter and flat space, the largest entangling region gives a thermal density matrix --Rindler in flat space, and the static patch in dS.

\section{Final remarks and future directions}\label{sec:concl}

In this work, we used strong subadditivity of the EE and the Markov property of the CFT vacuum to establish the irreversibility of the renormalization group for general de Sitter invariant QFTs in $d=2$, $3$ and $4$ dimensions. Our results show the promising role of quantum information theory methods in cosmology, and we would like to end by discussing some future directions.

It would be interesting to evaluate our irreversibility formula in free theories in de Sitter. Even in flat space this requires numerical lattice calculations \cite{Casini:2009sr}; these models turn out to be considerably more challenging in curved spacetime, since they require formulating a lattice in an expanding or contracting spacetime (see e.g. \cite{Foster:2004yc}). We are currently developing these lattice methods, obtaining explicit examples that illustrate the general theorems that we have presented here; this will be reported in future work. These explicit calculations would also allow to understand strong infrared effects due to light scalars in de Sitter from the point of view of quantum information. In this direction, the recent work \cite{Boutivas:2024lts} studied entanglement entropy for the massless scalar on the lattice in the inflationary patch, using an infrared length regulator $L$ to render the infrared divergences finite. In 3+1 dimensions they found a new term in the entanglement entropy for a region of radius $R$ which, after substracting the entanglement entropy of the conformally coupled scalar, gives $\Delta S(R)=\frac{1}{3} \frac{R^2}{\ell^2} \log(L/R)$. This UV finite term provides a nontrivial check for our irreversibility formula (\ref{eq:irrev2}) in $d=4$ space-time dimensions. A complementary approach would be to extend our framework to regions much larger than the Hubble scale, where analytic approximations are also possible \cite{Maldacena:2012xp}.

Another future direction that we think is worth pursuing is the study of quantum information measures for strongly coupled QFTs in de Sitter that admit a holographic description.  See e.g. \cite{Marolf:2010tg} for some examples. This would provide a framework for an independent gravitational analysis of dual RG flows and their irreversibility properties.  This would naturally be complementary with results on free theories.

In our approach, we kept the dS radius fixed and used the size of the entangling region $R$ to probe the RG, avoiding issues that can arise when relying on the thermal entropy instead \cite{Ben-Ami:2015zsa}. It would be interesting to work out the thermal interpretation of our results in the limit when $R \to \ell$, likely connecting with recent developments in cosmological bootstrap techniques \cite{Baumann:2022jpr}. We expect that light-ray operators and energy conditions may also play a fruitful role in de Sitter.  We are also currently analyzing the extension of our methods to QFT in Anti de Sitter spacetime \cite{Abate:2026apg}.

\section*{Acknowledgments} 

We thank H. Casini for discussions. 
NA is supported by a CONICET PhD fellowhip. GT is supported by CONICET (PIP grant 11220200101008CO), CNEA, and Instituto Balseiro, Universidad Nacional de Cuyo. We acknowledge support from the Simons Foundation by a targeted grant to Instituto Balseiro. 

\appendix

\section{Structure of the CFT vacuum EE on the light-cone}\label{appendix:K}

In this Appendix we will give a more explicit expression for the Markovian EE \eqref{eq:localS}. It should be constructed using geometric quantities associated to the metric $h_{ij}= \ell^2\,\tan^2 \gamma\, \hat g_{ij}$ of (\ref{eq:dgamma1}), plus a possible anomaly term. The available geometric objects are the metric, its Riemann tensor, and the extrinsic curvatures. Since the space is conformally flat, the Riemann tensor can be obtained from the Ricci tensor. Furthermore, we will now show that the extrinsic curvatures can also be constructed from the induced metric and the Ricci tensor. We will set $\ell=1$ throughout this section.

The 2-dimensional vector space normal to the surface is generated by two null vectors:
\begin{align}
n_\mu^{(1)}&=\delta_\mu^T+\delta_\mu^\theta\,,\\
n_\mu^{(2)}&=q_\mu-\frac{(\nabla\gamma)^2}{2\cos^4 T}\,n^{(1)}_\mu-\frac{1}{\cos^2 T}\nabla_\mu\gamma\,,
\end{align}
where $q^\mu=\frac{1}{2}(\delta^\mu_T+\delta^\mu_\theta)$ is another null vector that satisfies $q\cdot n^{(1)}=1$. These two covectors $n^{(1)}, n^{(2)}$ are normal to the tangent vectors,
\be
e_i^\mu=\frac{\partial x^\mu}{\partial \hat x^i}(-\partial_i\gamma,\partial_i\gamma,\delta_i^\mu)\,,
\ee
namely $n^{(1)}_\mu e_i^\mu=n^{(2)}_\mu e_i^\mu$.
Then, the extrinsic curvatures are
\be
K_{\mu\nu}^{(I)}=P_\mu^\alpha P_\nu^\beta\nabla_\alpha n^{(I)}_\beta\,,\quad I=1,2\,.
\ee
Here $P$ is the projector onto the tangent space to the curve, namely
\be
P_\mu^\nu=\delta_\mu^\nu-n^{(1)}_\mu n^{(2)\,\nu}-n^{(2)}_\mu n^{(1)\,\nu}\,.
\ee
We note that the induced metric on the surface is given by $h_{\mu\nu}=P_{\mu\nu}$, and its pullback on the tangent space is
\be\label{eq:h_pullback}
e^\mu_ie^\nu_jh_{\mu\nu}=h_{ij}=\tan^2\gamma(\hat x)\,\hat{g}_{ij}\,.
\ee

We want to show that both extrinsic curvatures on the surface can be constructed using geometric quantities associated with $h_{ij}$. After a few manipulations, one arrives at
\be\label{eq:K1}
K_{\mu\nu}^{(1)}=\frac{1}{\tan\gamma}\,h_{\mu\nu}\ \Rightarrow\ e^\mu_ie^\nu_jK_{\mu\nu}^{(1)}=\frac{1}{\tan\gamma}\,h_{ij}.
\ee
and
\begin{align}\label{eq:K2}
K_{ij}^{(2)}
&=-\frac{1}{2}\,(\tan\gamma-\cot\gamma)\, h_{ij}+\frac{(\nabla\gamma)^2}{2\cos^3\gamma\sin\gamma}\,h_{ij}\notag\\&\quad-\frac{\nabla_i^\mathrm{int}\nabla_j^\mathrm{int}\gamma}{\cos^2\gamma}-\frac{2\tan \gamma}{\cos^2\gamma} \nabla_i^\mathrm{int} \gamma\,\nabla_j^\mathrm{int} \gamma\,.
\end{align}
These formulas establish our claim. For $d>4$, using the relations between the Ricci tensors and curvatures associated with $h_{ij}$ (which we will denote as $\mc R_{ij}$ and $\mc R$) and $\hat{g}_{ij}$ \cite{Wald:1984rg}, one can rewrite \eqref{eq:K2} more compactly in terms of intrinsic geometric quantities as
\be
K_{ij}^{(2)}=\frac{\tan\gamma}{d-4}\of{\mathcal{R}_{ij}-\frac{1}{2(d-3)}\,h_{ij}\mathcal{R}}-\frac{\tan\gamma}{2}\,h_{ij}\,.
\ee
Furthermore, from the expressions \eqref{eq:K1} and \eqref{eq:K2}, one can check that the invariant associated with the type B anomaly in the universal logarithmic term of the EE in $d=4$ \cite{Solodukhin:2008dh} vanishes:
\be
K_{\mu\nu}^{(1)}K^{\mu\nu\,(2)}-\frac{1}{2}K^{\mu\,(1)}_\mu K^{\mu\,(2)}_\mu=0\,.
\ee

Let us now describe in more detail the Markovian expression for the CFT entanglement entropy.
For odd $d$ there are no anomalous terms, and the most general expression for the local function $L$ is
\be
L=\frac{\alpha_0}{\epsilon^{d-2}}+\frac{\alpha_2\,\mc R}{\epsilon^{d-4}} +\frac{\alpha_4\,{\mc R}^2}{\epsilon^{d-6}} +\frac{\alpha_4'\,{\mc R}_{ij}^2}{\epsilon^{d-6}} +\ldots
\ee

On the other hand, for even $d$ we have to include a contribution that accounts for the Euler type A anomaly. Following \cite{Casini:2018kzx}, we can achieve this by including a Wess-Zumino term for the `dilaton field' $\log(\tan\gamma(\hat{x})/\epsilon)$. In $d=4$ this reads
\be
S_\mathrm{WZ}^{(d=4)}[\gamma]=-4A\int\mathrm{d}\Omega\off{\log\of{\frac{\tan\gamma}{\epsilon}}+\frac{2(\nabla\gamma)^2}{\sin^2(2\gamma)}},
\ee
where the prefactor is fixed by the anomaly contribution for a sphere. Note that this term is a local functional of $\gamma$ and $\hat g_{ij}$, but not in terms of $h_{ij}$.

\bibliography{EE}{}
\bibliographystyle{utphys}

\end{document}